\documentclass[12pt,english,aps,prl,twocolumn,showpacs,superscriptaddress,groupedaddress,footinbib,reprint,noshowpacs]{revtex4-1}
\usepackage{graphicx}
\usepackage{amsmath}
\usepackage{comment}
\usepackage{amssymb}

\usepackage{subfigure}

\usepackage{subfigure}
\usepackage{color}
\usepackage{soul}

\begin{document}

\title{Inverse Ising inference by combining Ornstein-Zernike theory with deep learning}

\author{Soma Turi}
\affiliation{Cavendish Laboratory, University of Cambridge, Cambridge CB3 0HE, United Kingdom}

\author{Alpha A. Lee}
\email{aal44@cam.ac.uk}
\affiliation{Cavendish Laboratory, University of Cambridge, Cambridge CB3 0HE, United Kingdom}

\begin{abstract}
Inferring a generative model from data is a fundamental problem in machine learning. It is well-known that the Ising model is the maximum entropy model for binary variables which reproduces the sample mean and pairwise correlations. Learning the parameters of the Ising model from data is the challenge. We establish an analogy between the inverse Ising problem and the Ornstein-Zernike formalism in liquid state physics. Rather than analytically deriving the closure relation, we use a deep neural network to learn the closure from simulations of the Ising model. We show, using simulations as well as biochemical datasets, that the deep neural network model outperforms systematic field-theoretic expansions, is more data-efficient than the pseudolikelihood  method, and can generalize well beyond the parameter regime of the training data. The neural network is able to learn from synthetic data, which can be generated with relative ease, to give accurate predictions on real world datasets. 
\end{abstract} 

\makeatother
\maketitle

Drawing meaningful interference from correlations amongst variables is a fundamental problem in science. The central challenge is to decipher the probability distribution that generates the correlations given a set of observations, and then predict properties of unknown samples with the inferred model. Many probabilistic models have been proposed in the literature \cite{bengio2013representation,salakhutdinov2015learning}. Focusing on capturing the sample mean and pairwise correlations, the simplest model, in the sense of maximum entropy, is a Boltzmann probability distribution with a Hamiltonian that contains terms linear and bilinear in the variables \cite{jaynes1957information}. For continuous variables, this is the multivariate Gaussian distribution; for discrete variable, this model is the inverse problem of the Ising model in statistical physics. 

The inverse Ising model, also known as Boltzmann learning \cite{ackley1985learning}, has found applications in different disciplines \cite{nguyen2017inverse} such as understanding neural spike trains \cite{schneidman2006weak,cocco2009neuronal}, bird flocks \cite{bialek2012statistical}, and predicting structures of protein \cite{lapedes2012using,marks2011protein,morcos2011direct} and RNA \cite{de2015direct} as well as their fitness landscapes \cite{shekhar2013spin,ferguson2013translating,figliuzzi2015coevolutionary,hopf2015quantification} using evolutionary data.  The inverse Ising model disentangles correlations in the dataset into pairwise interactions, thus could distinguish direct variable-variable interactions from indirect correlations mediated through other variables. 

However, exact maximum likelihood inference of the inverse Ising model is numerically challenging because it requires computing the partition function at each step of the optimisation \cite{ackley1985learning}. To overcome this challenge, many approximate techniques have been developed  \cite{opper2001advanced}. Those techniques are typically leading terms of asymptotic expansions that relate the sample mean and correlation to the Ising parameters in limits where the coupling is tractable, e.g. asymptotically small correlations \cite{kappen1998efficient,tanaka1998mean,sessak2009small}, small number of coupled clusters \cite{cocco2011adaptive,cocco2012adaptive,barton2016ace}, a tree-like structure \cite{nguyen2012bethe}. (For comprehensive reviews, see \cite{cocco2017inverse,nguyen2017inverse}.)  A relatively recent class of algorithms maximize the pseudolikelihood rather than the likelihood  \cite{aurell2012inverse,decelle2014pseudolikelihood}. Although the pseudolikelihood method avoids partition function evaluations and converges to the maximum likelihood solution in the limit of infinite data, the algorithmic complexity scales with the number of samples. 

In this Letter, we eschew analytical asymptotic expansions and instead use a deep learning model to infer the relationship between the Ising parameters and sample mean and correlation. We will motivate an analogy between the inverse Ising model and the Ornstein-Zernike formalism in liquid state physics. We show that a deep neural network, trained using simulations of Ising models, can approximate an Ornstein-Zernike-like closure for the inverse Ising model. The deep neural network is generalizable and achieves an accuracy beyond analytical approximations for biochemical datasets as well as simulations with disparately different parameters compared to the training data. The neural network learns from synthetic data, which can be generated with relative ease, to give accurate predictions on real world datasets with a runtime independent of the dataset size.   

We begin by stating the inverse Ising model: We are given $p$ sequences of $N$ variables, $\{\mathbf{s}^{\alpha}\}_{\alpha=1}^{p}$, each variable $s_i$ can take values $\pm 1$. The maximum entropy distribution with the mean and pairwise correlations agreeing with the data, i.e.
\begin{equation}
\left< \sigma_i\right>_{\mathcal{P}} = \left< s_i\right>_{\mathrm{data}}, \quad \left< \sigma_i \sigma_j\right>_{\mathcal{P}} = \left< s_i s_j\right>_{\mathrm{data}}
\end{equation}
is the Ising model 
\begin{equation} 
\mathcal{P}(\boldsymbol{\sigma}) = \frac{1}{Z} \exp\left(  \sum_i h_i \sigma_i + \sum_{i>j} J_{ij}\sigma_i \sigma_j \right). 
\label{Boltzmann}
\end{equation} 
The inverse Ising problem is the inference of parameters $\left\{h_i, J_{ij} \right\}$  from data. This is challenging because the correlation between variables $i$ and $j$, $C_{ij} =\left<s_i s_j \right> - \left<s_i\right> \left<s_j\right>$, can be large even if $J_{ij} = 0$ if there is an intervening variable $k$ such that $J_{ik}$ and $J_{kj}$ are large. In other words, the pairwise correlations $C_{ij}$ one detect in a dataset is a measure of the interaction between $i$ and $j$ that is mediated by all other variables. 

To make further progress, we follow the Ornstein-Zernike formalism \cite{ornstein1914accidental,Hasen2013} in liquid state physics to deconvolve direct interactions and indirect correlations. We first consider a one-component, homogenous and isotropic liquid. Molecules interact with a pairwise additive potential $v(\mathbf{r}_{12})$, where $\mathbf{r}_{12}$ is the distance between particles 1 and 2. The liquid structure is characterised by the radial distribution function, $g(\mathbf{r}_{12})$, which is the probability of observing a molecule at distance $\mathbf{r}_{12}$ away from a molecule at the origin. Ornstein and Zernike noticed that $g(\mathbf{r}_{12})$ can be long ranged even when $v(\mathbf{r}_{12})$ is short ranged because $g(\mathbf{r}_{12})$ accounts for both the direct interaction between two molecules as well as the indirect interactions with surrounding molecules. Their crucial insight is to introduce a quantity known as the direct correlation function, $c(\mathbf{r}_{12})$, and write the total correlation function $h(\mathbf{r}_{12}) \equiv g(\mathbf{r}_{12})-1$ as 
\begin{equation}
h(\mathbf{r}_{12}) = c(\mathbf{r}_{12}) + \int \mathrm{d}\mathbf{r}_3 \; c(\mathbf{r}_{13}) c(\mathbf{r}_{32}) + \cdots 
\label{multipleint}
\end{equation}
where the first term captures the direct influence of molecule 1 on 2, the second term captures the influence of molecule 1 on 2 mediated by molecule 3, and higher order terms capture correlations induced by more intervening molecules. Equation (\ref{multipleint}) can be rewritten in a compact form $h(\mathbf{r}_{12}) =  c(\mathbf{r}_{12}) + \int \mathrm{d}\mathbf{r}_3 \; c(\mathbf{r}_{13}) h(\mathbf{r}_{32})$ which is known as the Ornstein-Zernike equation (we have scaled the standard correlation functions by the density to make the link with the Ising model more apparent later). To close the problem, we need a relation between $c(\mathbf{r}_{12})$, $h(\mathbf{r}_{12})$, and $v(\mathbf{r}_{12})$. The crucial feature of many closures is that they are \emph{local} and take the form 
\begin{equation}
f(c(\mathbf{r}_{12}),h(\mathbf{r}_{12}),v(\mathbf{r}_{12});\rho) = 0
\end{equation}
where $f$ is a real-valued function (\emph{not} a functional) and $\rho$ is the liquid density \cite{Hasen2013}. The locality of the closure relationship is approximate \cite{fantoni2004computer}, albeit a very good approximation for a wide variety of inter-particle interactions. 

Having introduced the Ornstein-Zernike formalism and closure, we return to the inverse Ising problem. The sample correlation, $C_{ij}$, is a discrete analogy of $h(\mathbf{r}_{ij})$ in the Ornstein-Zernike formalism. Therefore, we introduce the direct correlation $D_{ij}$ -- the discrete analogue of $c(\mathbf{r}_{ij})$ -- and replace the integrals in Equation (\ref{multipleint}) as a sum over lattice sites, i.e. 
\begin{equation}
C_{ij} = \delta_{ij}  + D_{ij} + \sum_k D_{ik} D_{kj} + \sum_{k,l} D_{ik} D_{kl} D_{lj} + \cdots. 
\end{equation} 
In matrix form, $C = (\mathbb{I} - D)^{-1}$ or 
\begin{equation}
D = \mathbb{I}  - C^{-1}. 
\end{equation} 
Taking $J = -C^{-1}$ is known as the mean-field approximation or Direct Coupling Analysis \cite{marks2011protein,kinjo2015liquid}. 

Now we introduce the key assumption of this Letter: we posit that the locality heuristic of closure relations for liquids applies to the inverse Ising problem. This crucial assumption will be verified later by comparing with simulations. We posit a function of four real variables $F(\cdot, \cdot,\cdot,\cdot)$ such that
\begin{equation}
J_{ij} \approx F\left(C_{ij}, [C^{-1}]_{ij},  \left<s_i\right>,  \left<s_j\right> \right). 
\label{close_J}
\end{equation}
Unlike the case of homogeneous fluids, we need an extra equation to determine the fields $h_i$. Motivated by the free energy of inhomogeneous fluids \cite{percus1988free}, we posit a function $G(\cdot, \cdot,\cdot,\cdot)$ such that
\begin{equation}
h_i \approx G\left( \tanh^{-1}\left<s_i\right>, [C^{-1}]_{ii}, \sum_{j\neq i} J_{ij} \left<s_j\right>, \sum_{j\neq i} C_{ij} \left<s_j\right> \right),  
\label{local_field}
\end{equation}
where $J_{ij}$ is given by $F$. We note that the first few terms of most approximate analytical inverse Ising theories do take the form of Equations (\ref{close_J})-(\ref{local_field}) with different $F$ and $G$ depending on the approximations  \cite{nguyen2017inverse}. 

$F$ and $G$ are complicated functions. Rather than analytically determining what they are, we will use a deep learning approach and approximate them using a richly parameterised interpolating function. Our hypothesis is that the neural network is able to find $F$ and $G$ that are better than mean-field expansions by inferring directly from data.


The training data is prepared by running 20 simulations with $N=50$ Ising sites. In each simulation, the Ising parameters are drawn from a normal distribution, $J_{ij} = J_{ji} \sim \mathcal{N}(0,\beta_J/\sqrt{N})$ and $h_i \sim \mathcal{N}(0,\beta_h)$, i.e. the Sherrington-Kirkpatrick model \cite{sherrington1975solvable} with a random field. The variances of the distributions are fixed in each simulation but vary across the 20 simulations, with $\beta_J \sim \mathrm{uniform}(0.5,2.5)$ and $\beta_h \sim \mathrm{uniform}(0.5,2.5)$. The one point and two point correlations are computed from $N \times 10^{7}$ steps of Markov Chain Monte Carlo (MCMC), sampled at every $10 N$ steps. 

$F$ and $G$ are approximated using multilayer neural networks. A $l$ layer neural network approximates functions by $l$ successive non-linear compositions, i.e. $y = \mathbf{W}_{l} \sigma(\mathbf{W}_{l-1} \cdots  \sigma(\mathbf{W}_1 \mathbf{x})))$, where $\mathbf{W}_{i} \in \mathbb{R}^{{M_{i} \times M_{i-1} }}$ is a weight matrix inferred from data, $M_{i}$ is the number of units in layer $i$, and $\sigma(\cdot)$ is a non-linear function. We use a neural network as it is a numerically tractable way of representing complex functions, and the parameters $\mathbf{W}_{i}$ can be efficiently inferred with backpropagation. The neural network architecture is discussed in the Supplemental Material and released on $\mathtt{github}$. 

The computational cost of evaluating the neural network to obtain one entry of  $J$ is independent of the number of variables or data, hence the total complexity is $O(N^2)$. The largest cost is computing $C^{-1}$ ($O(N^3)$). The pseudolikelihood approximation, the state of art method in the literature, has complexity $O(N^2 p)$ thus our algorithm is computationally less intensive than the pseudolikelihood method in the large data limit where the pseudolikelihood method is mathematically exact; we will show that our method is more data-efficient in the intermediate data regime.  



\begin{figure}
\includegraphics[scale=0.28]{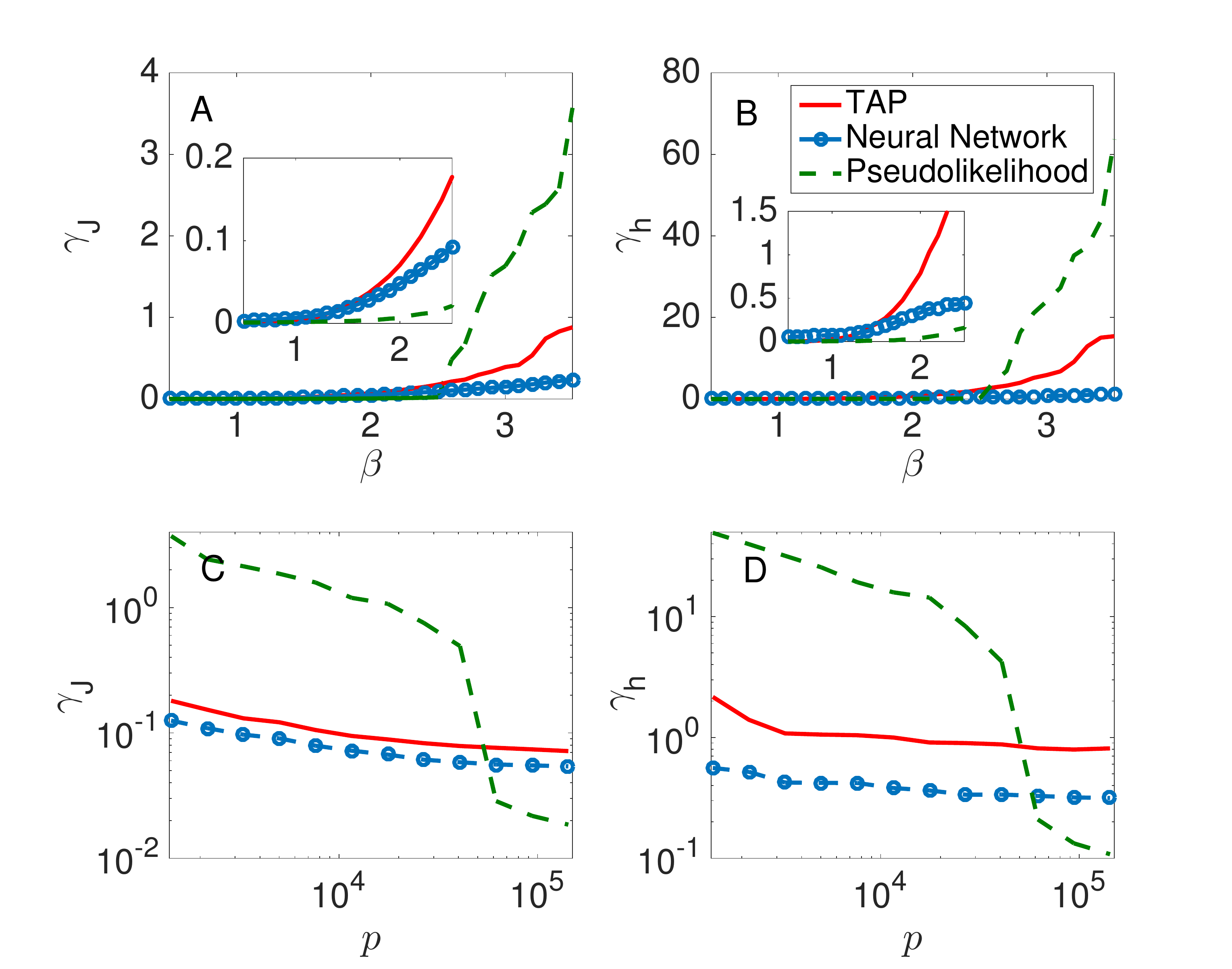}
\caption{The neural network model is accurate, generalisable and robust to sampling noise. The RMS error, $\gamma$, of predicting (A) $J_{ij}$ and (B) $h_i$ as a function of ``inverse temperature'' $\beta$. (C)-(D) The RMS error for predicting $J_{ij}$ and $h_i$ as a function of $p$, the number of MCMC samples used to estimate $C_{ij}$ and $\left<s_i\right>$, for $\beta = 2$.}
\label{generalise}
\end{figure}

To test the generality and accuracy of the model, we simulate Ising models with $N =70$ sites (note that the model is only trained on $N =50$ sites), $J_{ij} = J_{ji} \sim \mathcal{N}(0,\beta/\sqrt{N})$ and $h_i \sim \mathcal{N}(0,0.3\beta)$. A large value of $\beta$ corresponds to stronger coupling thus further away from the perturbative regimes that underlie analytical theories. Figure \ref{generalise}A-B shows the root mean square error of recovering $J_{ij}$ and $h_i$ as a function of $\beta$. The neural network model is more accurate than the Thouless-Anderson-Palmer approximation (TAP), a third order high temperature expansion \cite{thouless1977solution}. The pseudolikelihood approximation slightly outperforms that neural network at high temperature, where essentially all mean-field methods become accurate, but crucially the neural network outperforms the pseudolikelihood approximation at low temperatures where correlations become non-trivial.  Importantly, the neural network is accurate even for $\beta \in (2.5,3.5)$, which is outside the coupling strengths in the training data, demonstrating generalizability. The neural network approximation is also robust to sampling noise. Figure \ref{generalise}C-D shows that our neural network outperforms the TAP approximation as well as the pseudolikelihood approximation in the low data limit, demonstrating that our method is data efficient. 

\begin{figure}
\includegraphics[scale=0.3]{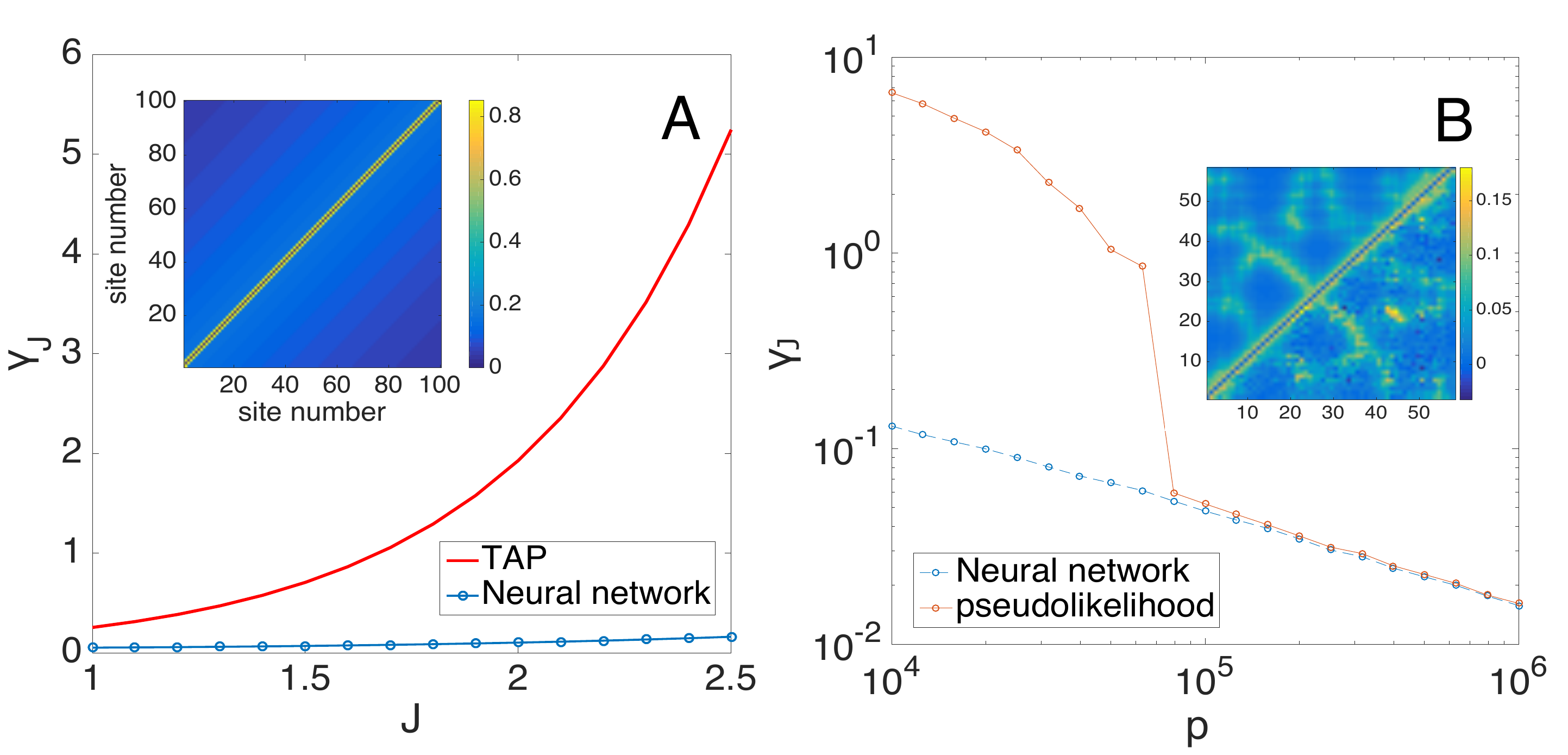}
\caption{The neural network model is accurate even for very non-Gaussian coupling matrices. (A) The RMS error, $\gamma_J$, of inferring $J$ for the nearest neighbour Ising model; the inset shows the inferred coupling matrix for $J=2.5$ . (B) Main panel: the RMS error against number of samples for the model protein contact map problem described in the main text. Inset: The top left corner shows the entries of the actual coupling matrix and the bottom right shows the entries of the inferred coupling matrix with $p = 10^{6}$ samples.}
\label{dist_coupling}
\end{figure}

The neural network approximation is generalizable and accurate even when the couplings are non-Gaussian distributed. Figure \ref{dist_coupling}A shows that the neural network approximation can accurately recover the coupling parameter for a 1D ferromagnetic Ising model with constant nearest-neighbour coupling $J_{ij} = J_{ji} = J (\delta_{i,j+1} + \delta_{i+1,j})$. The coupling matrix recovered from the neural network (inset of Figure \ref{dist_coupling}A) is strongly localised on the off-diagonal elements despite all the training data has a delocalised coupling matrix. We use the analytically determined correlation matrix as input to the neural network to focus on the error of the locality approximation, and Figure \ref{dist_coupling}A reassures that this intrinsic error is low. Inspired by the use of inverse Ising models in protein structure and fitness prediction \cite{marks2011protein,figliuzzi2015coevolutionary,hopf2015quantification}, Figure \ref{dist_coupling}B shows that the neural network approximation can accurately recover a model contact map, and is more data efficient compared to the pseudolikelihood method. In Figure \ref{dist_coupling}B, we consider the Bovine pancreatic trypsin inhibitor protein (PDB ID: 5PTI), a benchmark example in ref \cite{cocco2013principal} with $N=58$ amino acids. For this example, we assume the ``ground truth'' couplings are known (to what extent are amino acid interactions pairwise additive is a separate question \cite{jacquin2016benchmarking}), and take $J_{ij} = e^{-d_{ij}/7\AA}/\sqrt{N}$, where the $d_{ij}$ is the $C_\beta$ distance between residues $i$ and $j$. The correlation matrix is computed using MCMC; samples are taken every $10 N$ steps to ensure independence, and in total we acquire $p$ samples. The generalisability of the neural network to non-Gaussian distributed couplings confirms the approximate locality of the true Ornstein-Zernike-like closure. The functions $F$ and $G$ can be accurately approximated as long as the training data spans the four-dimensional input space. 

\begin{figure}
\includegraphics[scale=0.25]{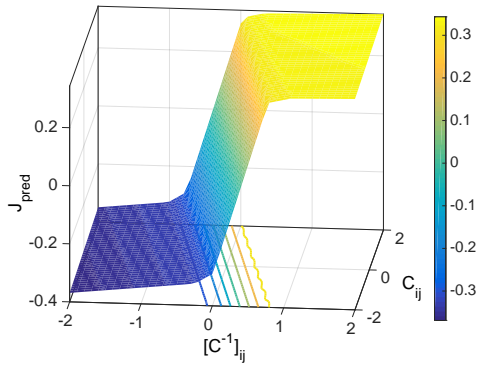}
\caption{The neural network learns to correct the overestimation of $J_{ij}$ in mean-field theory. The figure shows the coupling $J_{ij}$ predicted by the neural network closure as a function of $C_{ij}$ and $[C^{-1}]_{ij}$ for $\left<s_i\right>=0$. }
\label{interpretation}
\end{figure}

To interrogate what has the neural network learnt, we focus on the case $h_i =0$ and $\left<s_i\right>=0$ for all $i$ so that the closure is a 3D surface.  Figure \ref{interpretation} shows that the neural network learnt non-trivial corrections to pathologies in analytical theories: mean-field theory predicts $J_{ij} = - C^{-1}_{ij}$, yet this approximation significantly overestimates $J_{ij}$ \cite{barton2014large}. The neural network automatically corrects this by learning a sub-linear function to relate $J_{ij}$ to $C^{-1}_{ij}$. This is conceptually akin to phenomenologically imposing a large regularisation, a technique discussed in the physics literature \cite{barton2014large}, except the appropriate regularisation is inferred directly from data. Moreover, the neural network learns to use $C_{ij}$, another way to estimate the coupling, and only predicts a large value of $J_{ij}$ if both $C_{ij}$ and $C^{-1}_{ij}$ are large (c.f. the contours on the $C_{ij}$--$C^{-1}_{ij}$ plane). More generally, analytical approximate methods generally have some bias depending on aggregate quantities, e.g. inverse temperature and sparsity. The neural network learns those quantities by comparing $C$ and its inverse, and then applies an appropriate correction.

We will now go beyond synthetic data where the ground truth is known and apply our method to two problems in computational biology and chemistry. 

\begin{figure}
\centering
\includegraphics[scale=0.22]{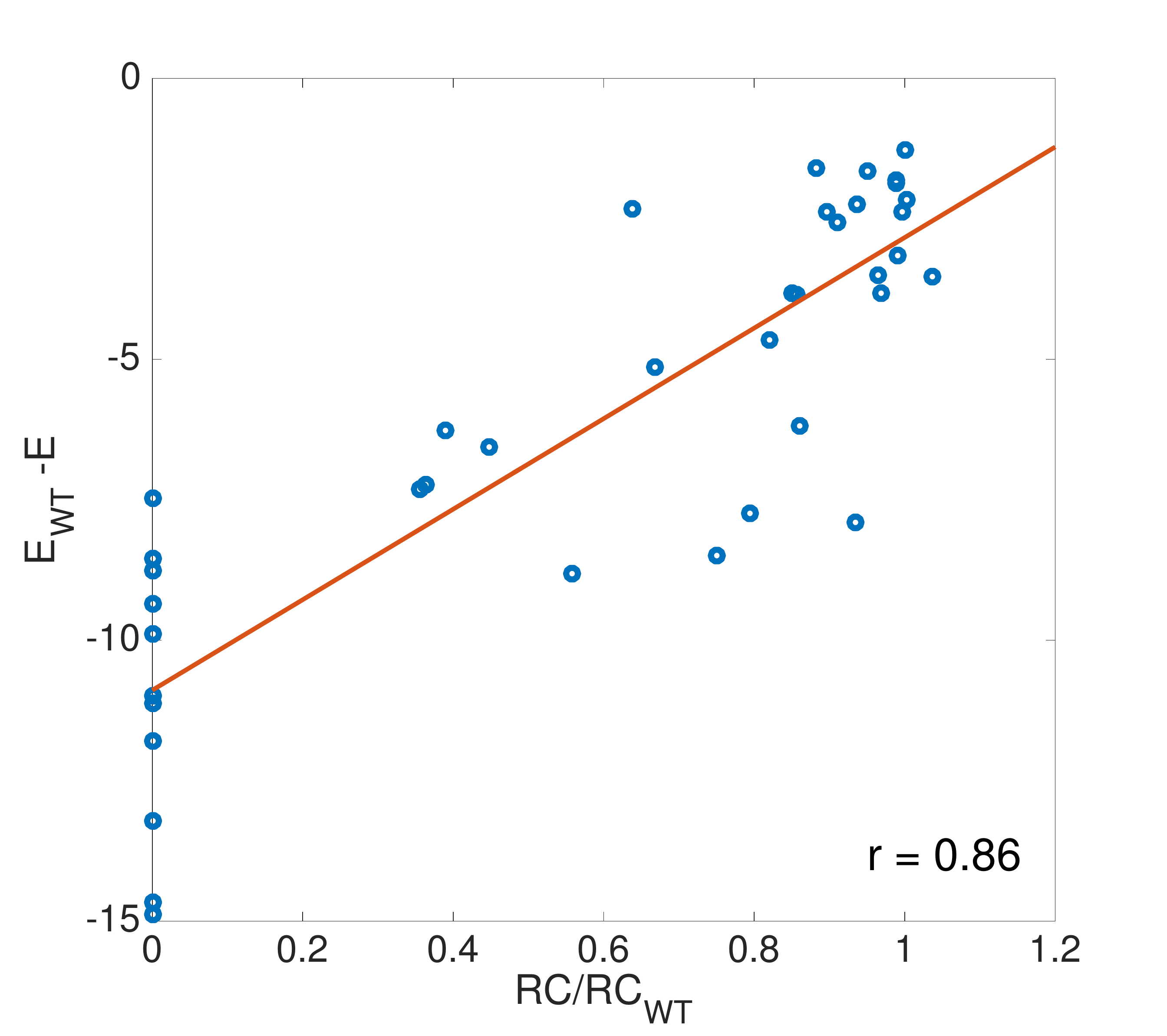}
\caption{The log probability of a mutant, $E$, is strongly correlated with the replication capacity $RC$. The subscript WT denotes wild type. Details of sequences and experiments can be found in ref \cite{mann2014fitness}. A pseudocount of $0.2$ is used to correct for undersampling \cite{durbin1998biological}, and the largest eigenvector of the correlation matrix, an artefact of amino acid conservation, is projected out before being processed by the neural network. }
\label{HIV}
\end{figure}

\emph{Fitness landscape of HIV-1 Gag}: Developing a predictive model for the fitness of HIV virus as a function of the amino acid sequence of the Group-specific antigen (Gag) protein is a challenge in vaccine development. Recent works showed that the fitness landscape can be inferred from the statistics of sequences found in patients \cite{ferguson2013translating}. The hypothesis is that the frequency of observing conserved sites and sets of correlated mutations reflect the contribution of those residues to fitness \cite{chakraborty2017rational}. Therefore, the fitness of an unknown sequence can be predicted by its log probability computed using a generative model for the sequences observed in patients. We replicate the analysis of ref \cite{mann2014fitness} for the HIV-1 Gag protein using the Ising representation for sequences, except the Ising parameters are inferred using the neural network. Figure \ref{HIV} shows that the log probability predicted by our model is highly correlated with experimental measurements of replication capacity, with a correlation coefficient $r = 0.86$, slightly higher than the state-of-the-art model \cite{mann2014fitness} ($r = 0.83$).  

\begin{figure}
\centering
\includegraphics[scale=0.22]{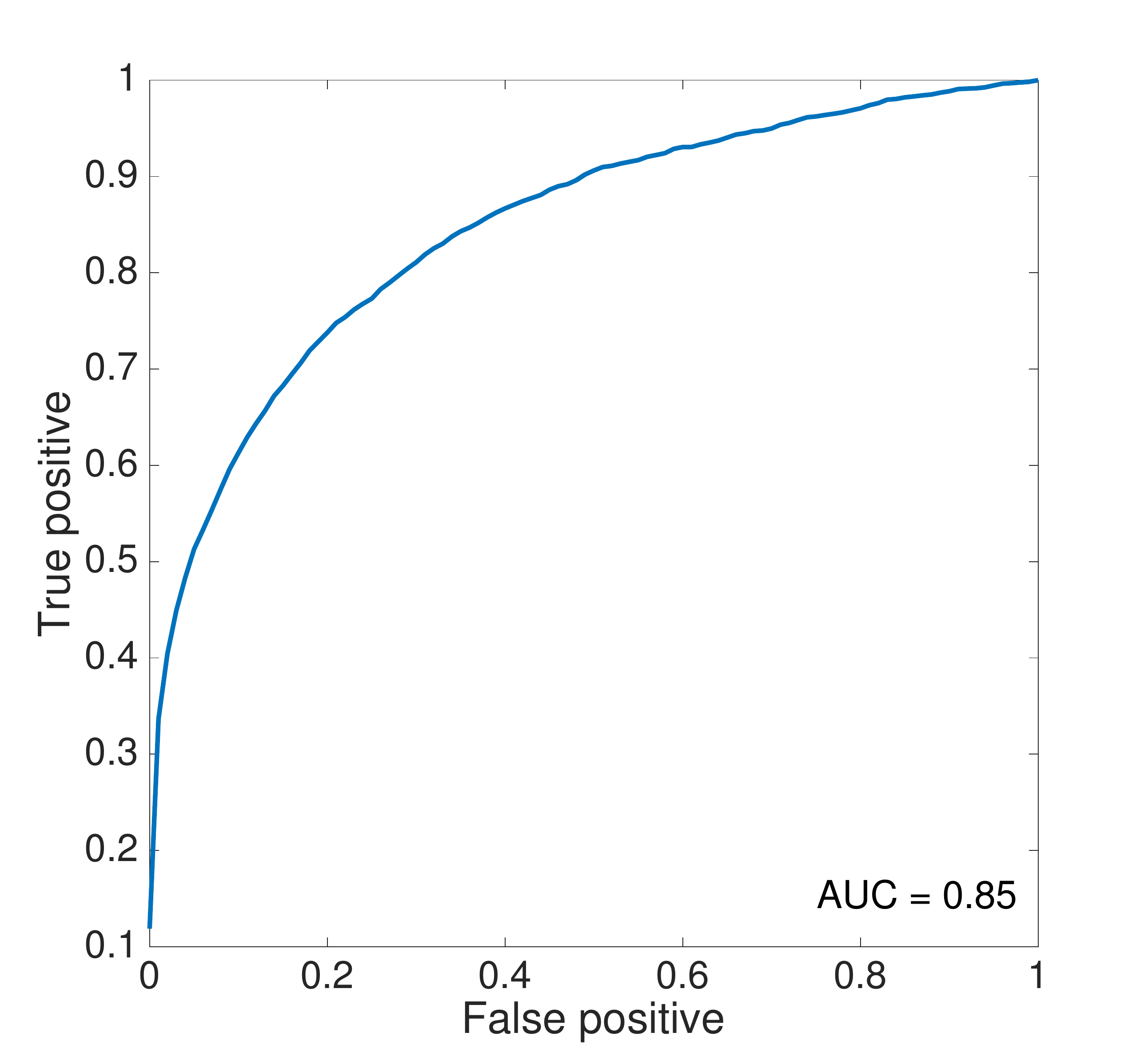}
\caption{The inverse Ising model accurately predicts protein-ligand binding for a panel of toxicologically relevant receptors in the Tox21 challenge \cite{huang2016modelling,huang2017editorial}. The figure shows the true positive rate averaged across the 12 receptors in the challenge as a function of false positive rate.}
\label{ER}
\end{figure}

\emph{Tox21 Challenge}: A key challenge in drug discovery is predicting whether an unknown molecule will bind to a particular receptor.  We consider the Tox21 challenge which reported binding affinities of small molecules against a panel of 12 toxicologically relevant receptors \cite{huang2016modelling,huang2017editorial}. Molecules are represented as a vector recording the presence (1)/absence (-1) of chemical groups (c.f. Supplemental Material). Each receptor is treated independently. As there are more variables compared to the number of data points, we undress finite sampling noise from the correlation matrix using an eigenvalue thresholding method inspired by random matrix theory  \cite{bouchaud2011financial,lee2016predicting,lee2017optimal} (c.f. Supplemental Material). Separate Ising models are inferred for active ($\{J^{\mathrm{b}}_{ij},h^{\mathrm{b}}_{i} \}$) and inactive ($\{J^{\mathrm{n-b}}_{ij},h^{\mathrm{n-b}}_{i} \}$) molecules. We score a molecule $\mathbf{f}$ by the log probability ratio $E(\mathbf{f}) = \sum_{i<j } f_i f_j (J^{\mathrm{b}}_{ij} - J^{\mathrm{n-b}}_{ij}) + \sum_{i } f_i (h^{\mathrm{b}}_{i} - h^{\mathrm{n-b}}_{i})$ and the molecule is predicted to bind if $E<\epsilon$, where $\epsilon$ controls the false/true positive tradeoff.  Figure \ref{ER} shows that the inverse Ising model accurately predicts binding (mean out-of-sample AUC across 12 receptors = 0.85). The state-of-the-art model achieves mean AUC of 0.83 but only when data from every receptor is pooled together \cite{wu2018moleculenet}, thus our model modestly outperforms the state-of-the-art, is approximately 12-times more data efficient, and clearly interpretable in terms of pairwise correlations between chemical features.   



In conclusion, we demonstrate a method that combines Ornstein-Zernike theory with a highly accurate closure parameterised using deep learning to solve the inverse Ising problem. We illustrate how our method can be used in real world datasets by considering examples in computational biology and chemoinformatics. We anticipate the strategy of parametrising Ornstein-Zernike closure with data to be applicable also in liquid state physics.

\begin{acknowledgments}
AAL thanks M. P. Brenner and R. Monasson for insightful discussions and comments, and J. P. Barton for making available the data of Ref \cite{mann2014fitness}. AAL acknowledges the Winton Programme for the Physics of Sustainability for funding. 

Data availability: Codes and data to reproduce results in this paper are available in \cite{git}. 
\end{acknowledgments}





\bibliography{references} 
\end{document}